\documentstyle[preprint,aps,epsfig]{revtex}
\begin{document}
\tightenlines
\draft
\lefthyphenmin=2
\righthyphenmin=3
\title{Limits on $WW\gamma$ and $WWZ$ Couplings from
$W$ Boson Pair Production}
\maketitle
\begin{center}
%
B.~Abbott,$^{31}$                                                             
M.~Abolins,$^{27}$                                                            
B.~S.~Acharya,$^{46}$                                                          
I.~Adam,$^{12}$                                                               
D.~L.~Adams,$^{40}$                                                            
M.~Adams,$^{17}$                                                              
S.~Ahn,$^{14}$                                                                
H.~Aihara,$^{23}$                                                             
G.~A.~Alves,$^{10}$                                                            
N.~Amos,$^{26}$
% ...
\\                                                                            
\vskip 0.50cm                                                                 
\centerline{(D\O\ Collaboration)}                                             
\vskip 0.50cm                                                                 
\sl{                                                                     
\centerline{$^{1}$Universidad de los Andes, Bogot\'{a}, Colombia}             
\centerline{$^{2}$University of Arizona, Tucson, Arizona 85721}               
\centerline{$^{3}$Boston University, Boston, Massachusetts 02215}             
\centerline{$^{4}$Brookhaven National Laboratory, Upton, New York 11973}      
\centerline{$^{5}$Brown University, Providence, Rhode Island 02912}           
\centerline{$^{6}$Universidad de Buenos Aires, Buenos Aires, Argentina}       
\centerline{$^{7}$University of California, Davis, California 95616}          
\centerline{$^{8}$University of California, Irvine, California 92697}         
\centerline{$^{9}$University of California, Riverside, California 92521}      
\centerline{$^{10}$LAFEX, Centro Brasileiro de Pesquisas F{\'\i}sicas,        
                  Rio de Janeiro, Brazil}                                     
\centerline{$^{11}$CINVESTAV, Mexico City, Mexico}                            
\centerline{$^{12}$Columbia University, New York, New York 10027}             
\centerline{$^{13}$Delhi University, Delhi, India 110007}                     
\centerline{$^{14}$Fermi National Accelerator Laboratory, Batavia,            
                   Illinois 60510}                                            
\centerline{$^{15}$Florida State University, Tallahassee, Florida 32306}      
\centerline{$^{16}$University of Hawaii, Honolulu, Hawaii 96822}              
\centerline{$^{17}$University of Illinois at Chicago, Chicago,                
                   Illinois 60607}                                            
\centerline{$^{18}$Indiana University, Bloomington, Indiana 47405}            
\centerline{$^{19}$Iowa State University, Ames, Iowa 50011}                   
\centerline{$^{20}$Korea University, Seoul, Korea}                            
\centerline{$^{21}$Kyungsung University, Pusan, Korea}                        
\centerline{$^{22}$Institute of Nuclear Physics, Krak\'ow, Poland}            
\centerline{$^{23}$Lawrence Berkeley National Laboratory and University of    
                   California, Berkeley, California 94720}                    
\centerline{$^{24}$Louisiana Tech University, Ruston, Louisiana 71272}        
\centerline{$^{25}$University of Maryland, College Park, Maryland 20742}      
\centerline{$^{26}$University of Michigan, Ann Arbor, Michigan 48109}         
\centerline{$^{27}$Michigan State University, East Lansing, Michigan 48824}   
\centerline{$^{28}$Institute for Theoretical and Experimental Physics,        
                   Moscow, Russia}                                            
\centerline{$^{29}$Moscow State University, Moscow, Russia}                   
\centerline{$^{30}$University of Nebraska, Lincoln, Nebraska 68588}           
\centerline{$^{31}$New York University, New York, New York 10003}             
\centerline{$^{32}$Northeastern University, Boston, Massachusetts 02115}      
\centerline{$^{33}$Northern Illinois University, DeKalb, Illinois 60115}      
\centerline{$^{34}$Northwestern University, Evanston, Illinois 60208}         
\centerline{$^{35}$University of Notre Dame, Notre Dame, Indiana 46556}       
\centerline{$^{36}$University of Oklahoma, Norman, Oklahoma 73019}            
\centerline{$^{37}$University of Panjab, Chandigarh 16-00-14, India}          
\centerline{$^{38}$Institute for High Energy Physics, Protvino 142284,        
                   Russia}                                                    
\centerline{$^{39}$Purdue University, West Lafayette, Indiana 47907}          
\centerline{$^{40}$Rice University, Houston, Texas 77005}                     
\centerline{$^{41}$Universidade do Estado do Rio de Janeiro, Brazil}          
\centerline{$^{42}$University of Rochester, Rochester, New York 14627}        
\centerline{$^{43}$CEA, DAPNIA/Service de Physique des Particules,            
                   CE-SACLAY, Gif-sur-Yvette, France}                         
\centerline{$^{44}$Seoul National University, Seoul, Korea}                   
\centerline{$^{45}$State University of New York, Stony Brook,                 
                   New York 11794}                                            
\centerline{$^{46}$Tata Institute of Fundamental Research,                    
                   Colaba, Mumbai 400005, India}                              
\centerline{$^{47}$University of Texas, Arlington, Texas 76019}               
\centerline{$^{48}$Texas A\&M University, College Station, Texas 77843}       
}                                                                             
\end{center} 


\begin{abstract}
The results of a  search for $W$ boson pair production in $p\bar{p}$ 
collisions at $\sqrt{s}=1.8$ TeV 
with subsequent decay to $e\mu$, $ee$, and $\mu\mu$  channels  
are presented.  
Five candidate events are observed 
with an expected background of $3.1\pm0.4$ events
for an integrated luminosity of approximately 97 pb$^{-1}$.
Limits on the anomalous couplings are obtained from a maximum likelihood fit
of the $E_T$ spectra of the leptons in the candidate events.
Assuming identical $WW\gamma$ and $WWZ$ couplings,
the $95 \%$ C.L. limits are $-0.62<\Delta\kappa<0.77 ~(\lambda = 0)$
and $-0.53<\lambda<0.56 ~(\Delta\kappa = 0)$
for a form factor scale $\Lambda = 1.5$ TeV.
\end{abstract}

\pacs{PACS numbers: 14.70.Fm 13.40.Em 13.40.Gp 13.85.Rm }


Gauge boson self-interactions are a direct consequence of 
the non-Abelian $SU(2)\times U(1)$ gauge symmetry of the standard 
model (SM). 
The trilinear gauge boson coupling strengths can be
measured directly by studying
gauge boson pair production.
Hadron collider experiments have established
the electroweak coupling of the $W$ boson to the photon~\cite{D0Wg} and
the existence
of the coupling between the $W$ boson and the $Z$ 
boson~\cite{CDFWWWZ,D0WWWZ}, and have placed constraints on anomalous 
$WW\gamma$ and $WWZ$ couplings~\cite{UA2,CDFWg,D01AWW,CDFWW,D0PRD}. 
Measurements of the couplings also have been reported by the
LEP collaborations~\cite{LEP}.

The $WW\gamma$ and $WWZ$ vertices can be described by a
general effective Lagrangian~\cite{Lagrangian}
with two overall coupling constants $g_{WW\gamma} = -e$ and
$g_{WWZ} = -e \cdot \cot \theta_{W}$ (where $e$ is the $W^+$ charge and
$\theta_{W}$ is the weak mixing angle)
and six dimensionless coupling parameters
$g_{1}^{V}$, $\kappa_V$, and $\lambda_V$ ($V = \gamma$ or $Z$),
after imposing {\it C}, {\it P}, and {\it CP} invariance.
Electromagnetic gauge invariance requires that $g_{1}^{\gamma} = 1$.
The effective Lagrangian becomes that of the {SM} when $g_1^{\gamma} = g_1^Z = 1$,
 $\kappa_{V} = 1 (\Delta\kappa_{V} \equiv \kappa_V - 1 = 0)$, and
 $\lambda_V = 0$.
In order to preserve unitarity at high energies, 
the anomalous couplings are modified by form factors with a scale $\Lambda$
(e.g. $\lambda_{V}(\hat{s}) = 
\lambda_{V}(0) / ( 1 + \hat{s}/\Lambda^{2})^{2}$).
Limits on these coupling strengths are usually obtained under the assumption
that the $WW\gamma$ and $WWZ$ couplings are equal
($g_1^{\gamma} = g_1^Z = 1$,
$\Delta\kappa_{\gamma} = \Delta\kappa_{Z}$, and $\lambda_{\gamma} =
\lambda_{Z}$),
leaving two independent couplings to be determined. 
In another approach~\cite{HISZ},
the anomalous couplings are formulated in a framework that explicitly respects
the $SU(2)\times U(1)$ gauge invariance, but
contains more general terms than those of Ref.~\cite{Lagrangian}.
Comparison of the two formalisms leads to simple equations (HISZ relations)
that relate anomalous couplings in the general effective Lagrangian.

In this paper we present the results of a search for 
$p\bar{p} \rightarrow WW+X 
\rightarrow \ell\bar{\ell}'\bar{\nu}\nu '+X$ at $\sqrt{s} = 1.8$ TeV,
where $\ell, \ell' = e$ or $\mu$. 
Limits on anomalous $WW\gamma$ and $WWZ$ couplings are 
obtained for both the equal couplings and the HISZ relations
by a maximum likelihood fit of the observed 
two-dimensional spectra of lepton transverse energy $E_T$.
This method provides tighter limits on anomalous couplings than those from
the measurement of the cross section~\cite{D01AWW,CDFWW}.
The $WW\rightarrow\ell\bar{\ell}'\bar{\nu}\nu '$ channel has significantly
less background, albeit with a smaller branching ratio, 
and is more sensitive to $WW$ production with the SM couplings
than the $WW/WZ\rightarrow \ell\nu jj/\ell{\bar \ell}jj$ 
channel. Therefore, limits obtained from this analysis are complementary 
to those from the
$WW/WZ\rightarrow \ell\nu jj/\ell{\bar \ell}jj$ analyses~\cite{CDFWWWZ,D0WWWZ}.

The data sample  corresponds to an integrated luminosity of
approximately 97 pb$^{-1}$ 
collected with the  D\O\ detector 
during the 1992--93 and 1993--1995 Tevatron collider runs at Fermilab.
The results based on the 1992--1993 data sample
of approximately 14 pb$^{-1}$ were previously reported~\cite{D01AWW,D0PRD}.
This paper describes the analysis of the 1993--1995 data sample and gives
the combined results from the two analyses.

The D{\O} detector\cite{D0Detector} consists of three major components:
the calorimeter, tracking, and muon systems.  
A hermetic, compensating, uranium-liquid argon sampling calorimeter with 
fine transverse and  longitudinal  segmentation in projective towers  
measures energy out to $|\eta| \sim 4.0$, where $\eta$ is the pseudorapidity.
The central and forward drift chambers are used to identify charged tracks
for $|\eta|\leq 3.2.$  There is no central magnetic field.  
Muons in the central region are identified and their momenta measured
with three layers of 
proportional drift tubes (PDT's), one inside and two outside of 
magnetized iron toroids, providing coverage for  $|\eta|\leq 1.7$. In 
addition, scintillation  counters mounted on the outer layer of PDT's 
provide time information for muon identification and cosmic ray
rejection.

Event samples are obtained from triggers with the signature of leptonic
$W$ boson
decays.  The $ee$ and $e\mu$ samples are selected from events passing a trigger 
which requires an electromagnetic cluster with $E_{T} > 20$ GeV and
missing transverse energy $\hbox{$\rlap{\kern0.25em/}E_T$} > 15$ GeV.
The integrated luminosity for this sample is 82.3 $\pm$ 4.4 pb$^{-1}$.  The 
$\mu\mu$ sample is selected from events passing a trigger which requires
at least one muon track in 
the fiducial region of $|\eta|\leq 1.0$ with $p_{T} > 15$ GeV/$c$, 
energy deposition in the calorimeter consistent with the passage of a muon, 
and no hits due to cosmic ray muons in the scintillator located outside of the
muon chambers.  
The integrated luminosity for this sample is 65.2 $\pm$ 3.5 pb$^{-1}$.

Isolated electrons are identified using a likelihood function formed from 
four variables: the electromagnetic energy fraction of the calorimeter
cluster, the 
$\chi^{2}$ of longitudinal and transverse shower shapes compared to 
test-beam 
and Monte Carlo electrons, the ionization energy deposition $(dE/dx)$ 
in the  central tracking detector associated with the matching track, and 
the distance between the projected track position and the centroid of the 
energy cluster at the calorimeter.
This likelihood function is 
used to discriminate between electrons and photon conversions, photon showers
overlapped 
with a charged hadron track, and hadronic showers with large electromagnetic
content.  
For a given identification efficiency, this method provides a background
rejection 
2-3 times higher than a method that places requirements on individual
variables~\cite{PBthesis}.
In the central region, $|\eta|\le 1.1$, the
electron detection efficiency is ($59.9 \pm 0.8$)\%; in the forward
region, $1.5 \le |\eta| \le 2.5$, it is ($47.1 \pm 1.4$)\%.

Muons are required to have associated hits in at least 
two of the three layers of the muon system.
They must be isolated from jets
($\Delta{\cal R}(\mu,{\rm jet})>0.5$ for $E_T^{\rm jet} > 10 {\rm GeV}$, where
$\Delta{\cal R}(\mu,{\rm jet})$ is the separation between muon and jet in
$\eta-\phi$ space)
and have energy deposition in
the calorimeter consistent with a minimum ionizing particle.
The muon track is required to point to the primary event vertex within 25 cm in 
the plane transverse to the beam 
and to occur at a time, as measured by the PDT's, 
within 200 ns of the beam crossing.
The muon detection efficiency is $(70.1\pm3.1)\%$ within
the fiducial acceptance of $|\eta|\leq 1.0$ employed in this analysis.

The $ee$ candidate events are selected by requiring the leading electron 
to have $E_{T} \geq 25$ GeV and a second electron to have 
$E_{T} \geq 20$ GeV.  The $\hbox{$\rlap{\kern0.25em/}E_T$}$
is required to exceed 25 GeV.
Background 
from events with high-$p_{T}$ $Z$ bosons is reduced by removing events with 
dielectron invariant mass within 15 GeV/$c^{2}$ of the $Z$ boson mass.
The $e\mu$ candidate events are selected by requiring an electron with 
$E_{T} \geq 25$ GeV and a muon with $p_{T} \geq 15$ GeV/$c$.
We require $\hbox{$\rlap{\kern0.25em/}E_T^{{\rm cal}}$}\geq 25$ GeV  
(where $\hbox{$\rlap{\kern0.25em/}E_T^{{\rm cal}}$}$
is $\hbox{$\rlap{\kern0.25em/}E_T$}$ calculated using
only the calorimeter and is not affected by the muon momentum resolution) and
$\hbox{$\rlap{\kern0.25em/}E_T$}\geq 20$ GeV.  
An isolation condition,
$\Delta {\cal R}(e,\mu) \geq 0.5$, where $\Delta {\cal R}(e,\mu)$ is the
separation between
electron and muon in $\eta$-$\phi$ space, 
is applied to remove background from cosmic ray muons accompanied by a
bremsstrahlung photon.
The $\mu\mu$ candidate events are selected by requiring a leading muon with 
$p_{T} \geq 25$ GeV/$c$ and a second muon with $p_{T} \geq 20$ GeV/$c$.  The 
projection of $\hbox{$\rlap{\kern0.25em/}E_T$}$
onto the bisector of the muon tracks in the 
transverse plane is required to exceed 30 GeV in order to remove background 
from $Z$ boson production.  This is less sensitive to the muon 
momentum resolution than a dimuon invariant mass requirement.

Additional cuts are applied similarly to all three channels. To remove
background from $Z \rightarrow \tau \tau$, and from $b{\bar b}$
production in the $\mu\mu$ channel, the transverse opening angle
$\Delta\phi_\ell$ between one charged lepton $\ell$ and
\hbox{$\rlap{\kern0.25em/}E_T$} is required to be
less than 160$^\circ$.  This cut is applied to the second-leading electron
in the $ee$ channel, the muon in the $e\mu$ channel, and the leading muon
in the $\mu\mu$ channel.  In addition, for the $ee$ and $e\mu$ channels,
$\Delta\phi_\ell$ is required to exceed 20$^\circ$, and both
$\Delta\phi_\ell$ requirements are removed if
$\hbox{$\rlap{\kern0.25em/}E_T$} > 50$ GeV.  Also, to reduce
background in all three channels from $t {\bar t}$ production, the
hadronic $E_T$ in the event,
$\vec{E}_{T}^{\rm had}\equiv 
-(\vec{E}_{T}^{\ell 1}+\vec{E}_{T}^{\ell 2}
+\vec{\hbox{$\rlap{\kern0.25em/}E_T$}})$, 
is required to satisfy $E_T^{\rm had} <
40$ GeV.  After imposing these selection criteria, one $ee$ candidate,
two $e\mu$ candidates, and one $\mu\mu$ candidate remain.

The detection efficiencies for $W$ boson pair production with SM and
anomalous couplings are
determined using a fast Monte Carlo program (the Monte Carlo event generator
of Ref.~\cite{Zeppenfeld} plus
a parametric  detector simulation).
The detection efficiencies for SM $W$ boson pair production are also calculated
using the {\small PYTHIA}~\cite{PYTHIA} 
event generator followed by a detailed  {\small GEANT}~\cite{GEANT}
simulation of the D\O\ detector and 
are found to agree with those
determined from the fast Monte Carlo.
Trigger and particle identification efficiencies 
are determined from the data.
The trigger efficiency for the $ee$ and $e\mu$ data samples is
$(99^{+1}_{-3})\%$.
For the $\mu\mu$ sample, the trigger efficiency is $(68.7\pm5.8)\%$.
Table~\ref{table:eff} shows the detection efficiencies for SM
$W$ boson pair production events and the 
number of expected events based on
a cross section of 9.4 pb~\cite{Ohnemus}.
The systematic uncertainty in the detection efficiency 
comes from electron (2.2\%) and muon (7.5\%) identification,
electron (2.0\%) and muon (8.5\%) trigger efficiencies, and the difference
between the detection efficiencies estimated with the two Monte Carlo methods
(5\%).
Uncertainty due to the choice of parton distribution 
function and evolution scale (5\%) is included in the uncertainty on the
number of expected events.

Backgrounds due to Drell-Yan dileptons, $W\gamma$, $t\bar{t}$, and $Z$ 
boson production are estimated using the {\small PYTHIA},
{\small ISAJET}~\cite{ISAJET} and  {\small HERWIG}~\cite{HERWIG}
Monte Carlo event generators, followed by the detailed
{\small GEANT} simulation of the D{\O} detector. 
Backgrounds due to high-$p_{T}$ $Z\rightarrow ee$ and 
$\mu\mu$ events are studied using a Monte Carlo event
generator based on the theoretical model of Ref.~\cite{Arnold}
and the parametric detector simulation.
Backgrounds from multijet and $W$ + jet events  with a jet misidentified as an
electron and with heavy quark production of isolated muons 
are estimated from the data.  
The probabilities for a jet to be misidentified as an electron and
for a jet to be accompanied by a muon that satisfies the isolation criterion
are measured from a large sample 
of events passing jet triggers.  Events with large
$\hbox{$\rlap{\kern0.25em/}E_T$}$ are rejected from 
this sample to remove
$W$+jets events.
For electrons, the misidentification probability is found to be
a slowly rising linear function of jet $E_T$
($3.7\times10^{-5}$ at 20 GeV, $1.9\times10^{-4}$ at 100 GeV in the central
region; and $3.5\times10^{-5}$ at 20 GeV, $1.8\times10^{-4}$ at 100 GeV in the
forward region),
while for muons it is found to be constant
($1.5\times10^{-5}$ for the $e\mu$ sample and $1.5\times10^{-4}$
for the $\mu\mu$ sample).
The background estimates are 
summarized in Table~\ref{table:BG}.
Systematic uncertainties include those
listed above 
as well as the uncertainty on the production cross section of the background
processes.

The number of candidate events, four in the 1993--1995 data sample (five when
the 1992--1993 data sample with one $ee$ candidate and $0.6 \pm 0.1$
background events is included), is consistent with
an expected SM $WW$ signal of $1.5\pm0.1$ ($1.9\pm0.1$) events plus
an estimated background of $2.5\pm0.4$ ($3.1\pm0.4$) events.
We have studied the stability of the results by relaxing some of the event
selection requirements, e.g. $E_T^{\rm had}$ and dielectron invariant mass
criteria.
The increases in the numbers of candidate and background events are found to
be consistent with the expectations.
An upper limit on the $W$ boson pair production cross section is calculated
from the number of the candidate events and the estimated background events
using the Poisson-distributed number of events convoluted with Gaussian
uncertainties on the detection efficiencies, background, and luminosity.
For SM $W$ boson pair production, the upper limit for the cross section
is 37.1 pb at the $95 \%$ C.L. using the 1992--1993 and 1993--1995
data samples.
The probability that the observed number of events correspond to a fluctuation
of the background, with no signal, is $20.6 \%$.

By studying $E_T$ spectra of leptons from $W$ boson pair candidates,
limits can be obtained on the anomalous $WW\gamma$ and $WWZ$
couplings.
Use of this kinematic information provides significantly tighter constraints on 
anomalous couplings than those from the measurement of the cross section
(the method 
used in previous $WW\rightarrow \ell\bar{\ell}'\bar{\nu}\nu '$ 
analyses~\cite{D01AWW,CDFWW}),
since the predicted increase in the gauge boson pair production cross section
with anomalous couplings is greater at higher gauge boson $p_T$.
A binned maximum likelihood fit is performed to the measured
spectra of $E_T$ of the two leptons in the event. 
Two-dimensional bins in $E_T$ of one lepton versus
$E_T$ of the other lepton are used in order to take into
account the correlation between the two leptons in the event.
The binnings used in the fit are shown in Table~\ref{table:bins}.
The probability for the sum of the background estimate and Monte Carlo
$WW$ signal prediction to fluctuate to the observed number of
events is calculated in each bin for a given set of anomalous couplings.
The uncertainties on the background estimates, efficiencies, integrated
luminosity, and theoretical prediction of the $WW$ production cross section are
convoluted with Gaussian distributions into the likelihood function.
The likelihood functions are calculated for
the 1992--1993 and 1993--1995 data samples separately and are combined taking
into account correlated uncertainties, such as theoretical uncertainties.

\begin{figure}[htb]
\centerline{\hbox{
\epsfig{figure=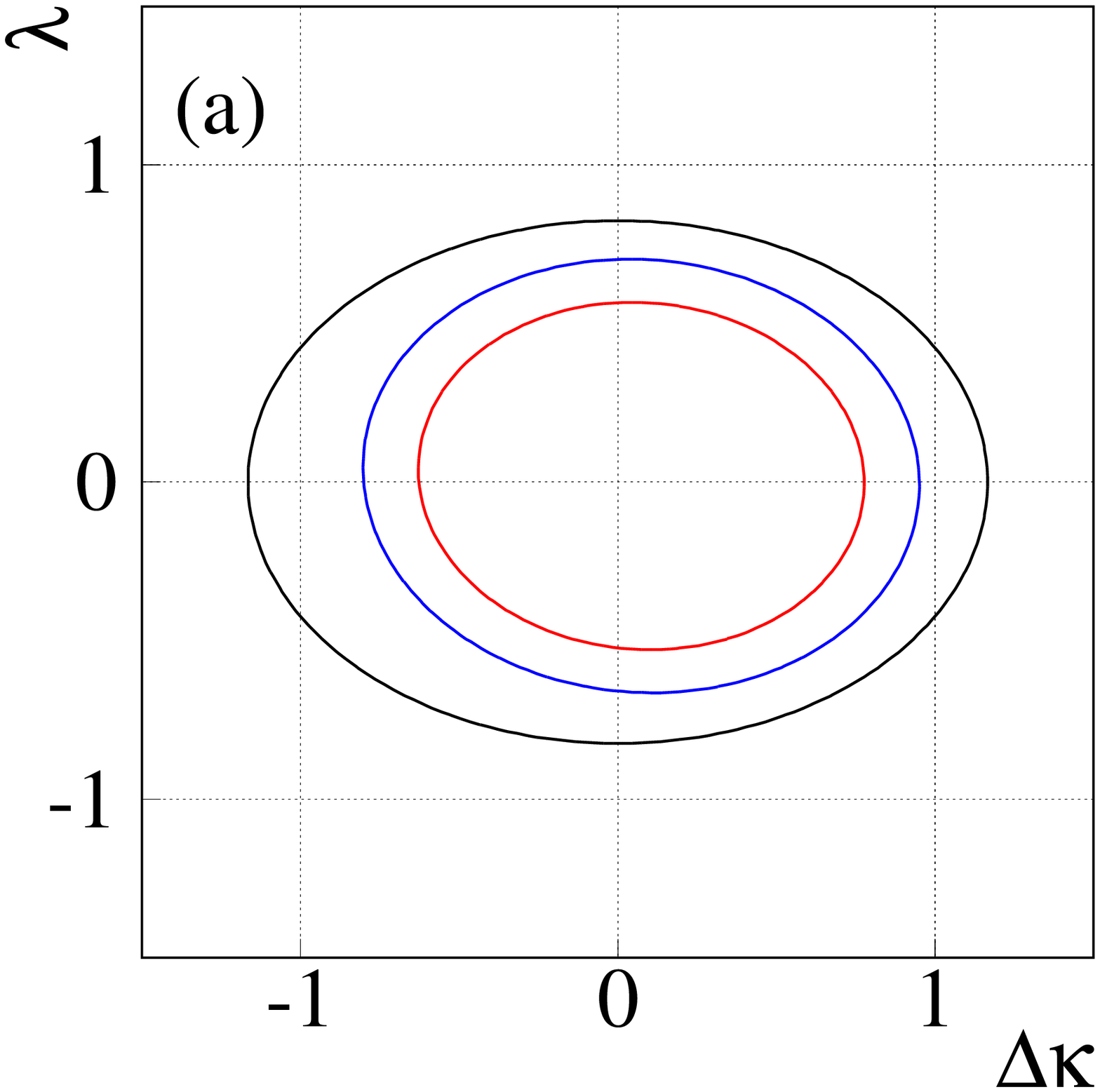,width=1.8in}
\epsfig{figure=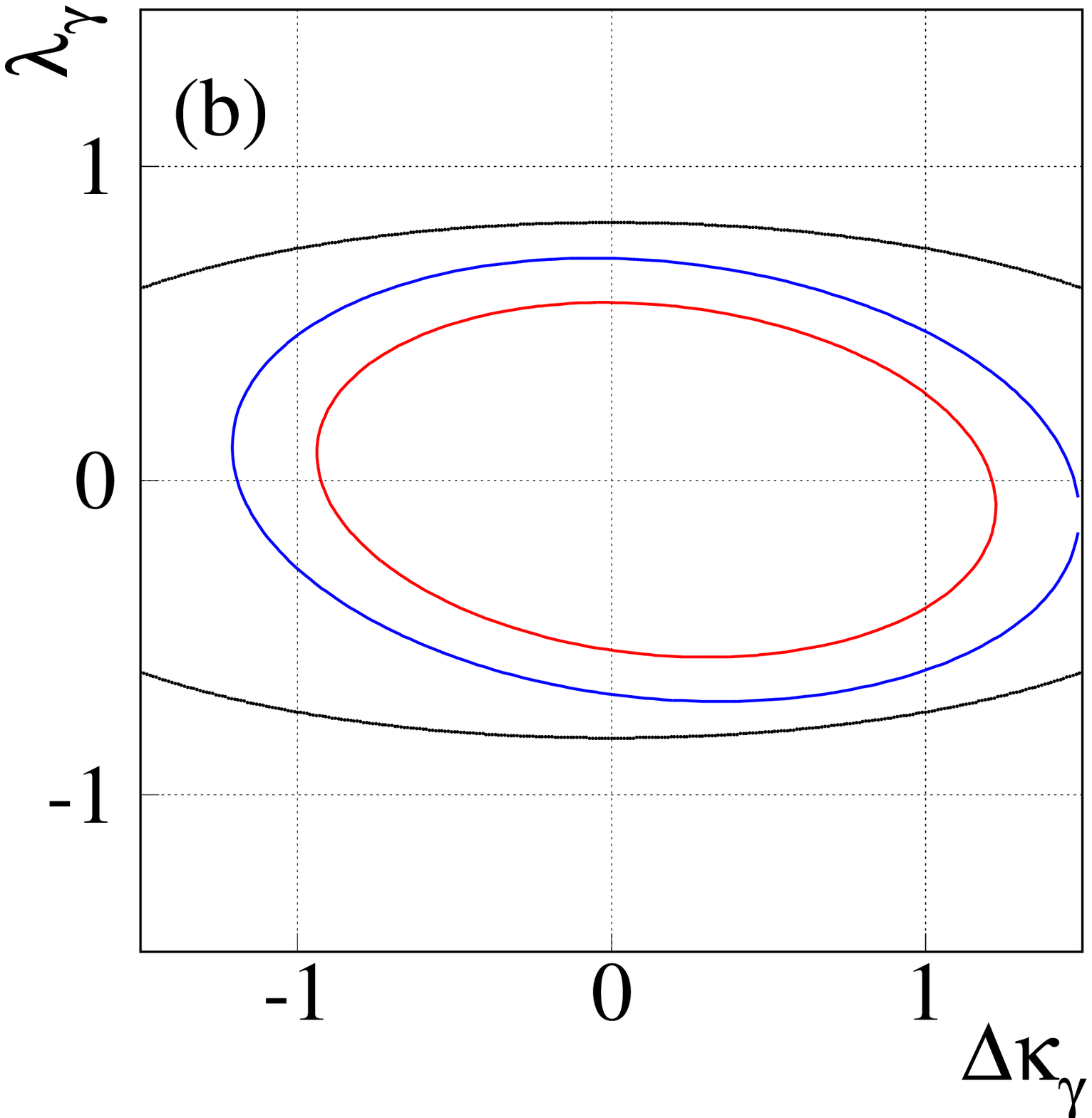,width=1.8in}}}
 \caption
 {Contour limits on anomalous couplings for $\Lambda = 1.5$ TeV:
 (a) $\Delta\kappa\equiv\Delta\kappa_\gamma=\Delta\kappa_Z$,
 $\lambda\equiv\lambda_\gamma=\lambda_Z$; and (b) HISZ relations.
 The innermost and middle curves are
 $95 \%$ C.L. one- and two-degree of freedom exclusion contours
 from the fit of the $E_T$ spectra of leptons, respectively. 
 The outermost curve is the constraint from the unitarity
 condition. Monte Carlo statistics limit the accuracy of the contours to
 $\pm0.01$.}
 \label{fig:contour}
\end{figure}

The $WW$ production process involves the $WW\gamma$ and $WWZ$ couplings,
unlike the $W\gamma$ production process which depends only on the $WW\gamma$
couplings.
Limits on anomalous couplings are obtained using two assumptions on the
relationship between the $WW\gamma$ and $WWZ$ couplings.
Figure~\ref{fig:contour} shows bounds on anomalous couplings from this
analysis and from the unitarity condition~\cite{Zeppenfeld,Unitarity}
using $\Lambda = 1.5$ TeV.
In Fig.~\ref{fig:contour}(a), the values for $\Delta\kappa$ and $\lambda$
are assumed to be equal for the $WW\gamma$ and $WWZ$ couplings.
Limits at the $95 \%$ C.L., when $\lambda$ or $\Delta\kappa$ is
set to zero, are:
\begin{eqnarray*}
-0.62 < \Delta\kappa < 0.77 ~(\lambda = 0);\\
-0.53 < \lambda < 0.56 ~(\Delta\kappa = 0)
\end{eqnarray*}
In Fig.~\ref{fig:contour}(b), the HISZ relations~\cite{HISZ} are used.
Limits at the $95 \%$ C.L. using the HISZ relations are:
\begin{eqnarray*}
-0.92 < \Delta\kappa_\gamma < 1.20 ~(\lambda_\gamma = 0);\\
-0.53 < \lambda_\gamma < 0.56 ~(\Delta\kappa_\gamma = 0)
\end{eqnarray*}
The innermost curve is the 95\% C.L. contour when only one coupling is treated
as a free parameter (e.g., limits on the axes) while the middle curve is the
95\% C.L. contour when both of the couplings are free parameters. 
All of the limits obtained in this analysis are comparable to those obtained
from the analysis of $WW/WZ\rightarrow e\nu jj$ events~\cite{D0WWWZ}.

In summary, a search for $WW\rightarrow\ell\bar{\ell}'\bar{\nu}\nu '$
in $p\bar{p}$ collisions at
$\sqrt{s} = 1.8$ TeV is performed using the 1992--1993 and 1993--1995 data
samples.
In approximately $97~{\rm pb}^{-1}$ of data,
five candidate events are found with an estimated background of $3.1 \pm 0.4$
events. From the standard model, $1.9 \pm 0.1$ events are expected.
The number of observed events is consistent with the standard model prediction
plus background estimate.
The $95 \%$ C.L. limits on the anomalous couplings
$-0.62 < \Delta\kappa < 0.77 ~(\lambda = 0)$ and
$-0.53 < \lambda < 0.56 ~(\Delta\kappa = 0)$ are obtained from a binned
maximum likelihood fit of the $E_T$ spectra of leptons,
assuming equal $WW\gamma$ and $WWZ$ couplings.

%
We thank the staffs at Fermilab and collaborating institutions for their
contributions to this work, and acknowledge support from the 
Department of Energy and National Science Foundation (U.S.A.),  
Commissariat  \` a L'Energie Atomique (France), 
State Committee for Science and Technology and Ministry for Atomic 
   Energy (Russia),
CAPES and CNPq (Brazil),
Departments of Atomic Energy and Science and Education (India),
Colciencias (Colombia),
CONACyT (Mexico),
Ministry of Education and KOSEF (Korea),
and CONICET and UBACyT (Argentina).

\begin{table}[htb]
\caption{Detection efficiencies and SM signal event expectations for
the 1993--1995 data sample. The uncertainties include both statistical
and systematic contributions.}
\begin{tabular}{lccc}
Channel                  & $ee$            &    $e\mu$      &  $\mu\mu$       \\ \hline
Detection Efficiency (\%)& $6.03 \pm 0.36$ & $4.76 \pm 0.49$& $1.19 \pm 0.18$ \\
SM Expectation (events)  & $0.52 \pm 0.04$ & $0.86 \pm 0.10$& $0.09 \pm 0.01$ \\ 
\end{tabular}
\label{table:eff}
\end{table}

\begin{table}[htb]
\caption{Summary of backgrounds and candidates for the 1993--1995 data sample.
The units are number of
events in the data sample. The uncertainties include both statistical
and systematic contributions.}
\begin{tabular}{lccc}
                                & $ee$            &  $e\mu$         
& $\mu\mu$        \\ \hline
Background:& & & \\
$Z\rightarrow$ $ee$ or $\mu\mu$ & $0.27 \pm 0.06$ &  $-$              
& $0.39 \pm 0.09$ \\
$Z\rightarrow \tau \tau$        & $0.10 \pm 0.07$ & $0.21 \pm 0.08$ 
& $<10^{-3}$      \\
Drell-Yan dileptons             & $0.03 \pm 0.04$ &  $-$                 
& $<10^{-3}$      \\
$W\gamma$                       & $0.18 \pm 0.07$ & $0.35 \pm 0.14$ 
& $-$             \\
$t\bar{t}$                      & $0.13 \pm 0.05$ & $0.18 \pm 0.06$ 
& $0.02 \pm 0.01$ \\ 
multijets/$W$ + jets            & $0.20 \pm 0.14$ & $0.43 \pm 0.28$ 
& $0.03 \pm 0.01$ \\
Total background                & $0.91 \pm 0.19$ & $1.17 \pm 0.33$ 
& $0.44 \pm 0.09$ \\
Data& 1& 2& 1\\
\end{tabular}
\label{table:BG}
\end{table}

\begin{table}[htb]
\caption{The binnings used in the maximum likelihood fit to set limits on
the anomalous couplings and the numbers of candidate events (background
estimate) for the 1992--1993 and 1993--1995 data samples.}
\label{table:bins}
\begin{center}
\begin{tabular}{ccc}
\multicolumn{3}{c}{$ee$ channel
 ($96.6\pm4.5~{\rm pb}^{-1}$)} \\ \hline
\multicolumn{1}{c|}{$E_{T}^{e1} ~\backslash ~E_{T}^{e2}$}&
\multicolumn{1}{c}{$20 - 40$ (GeV)}&
\multicolumn{1}{c} {$40 - 500$ (GeV)}\\ \hline
\multicolumn{1}{c|}{$25 - 40$ (GeV)}&
\multicolumn{1}{c}{2 ($0.50\pm0.10$)}&
\multicolumn{1}{c} {--}\\
\multicolumn{1}{c|}{$40 - 500$ (GeV)}&
\multicolumn{1}{c}{0 ($0.35\pm0.07$)}&
\multicolumn{1}{c} {0 ($0.27\pm0.06$)}\\ \hline\hline
\multicolumn{3}{c}{$e\mu$ channel
 ($96.2\pm4.5~{\rm pb}^{-1}$)} \\ \hline
\multicolumn{1}{c|}{$E_{T}^{e} ~\backslash ~E_{T}^{\mu}$}&
\multicolumn{1}{c}{$15 - 40$ (GeV)}&
\multicolumn{1}{c} {$40 - 500$ (GeV)}\\ \hline
\multicolumn{1}{c|}{$25 - 50$ (GeV)}&
\multicolumn{1}{c}{2 ($0.95\pm0.27$)}&
\multicolumn{1}{c} {0 ($0.16\pm0.05$)}\\
\multicolumn{1}{c|}{$50 - 500$ (GeV)}&
\multicolumn{1}{c}{0 ($0.16\pm0.05$)}&
\multicolumn{1}{c} {0 ($0.16\pm0.05$)}\\ \hline\hline
\multicolumn{3}{c}{$\mu\mu$ channel
 ($77.4\pm3.6~{\rm pb}^{-1}$)} \\ \hline
\multicolumn{1}{c|}{$E_{T}^{\mu 1} ~\backslash ~E_{T}^{\mu 2}$}&
\multicolumn{1}{c}{$20 - 40$ (GeV)}&
\multicolumn{1}{c} {$40 - 500$ (GeV)}\\ \hline
\multicolumn{1}{c|}{$25 - 40$ (GeV)}&
\multicolumn{1}{c}{1 ($0.08\pm0.02$)}&
\multicolumn{1}{c} {--}\\
\multicolumn{1}{c|}{$40 - 500$ (GeV)}&
\multicolumn{1}{c}{0 ($0.18\pm0.04$)}&
\multicolumn{1}{c} {0 ($0.26\pm0.05$)}\\
\end{tabular}
\end{center}
\end{table}

\end{document}